# Influence of control polygon on the generalization of the conversion between ANCF and B-spline surfaces

Peng Lan✉, Randi Wang, Zuqing Yu
School of Mechatronic Engineering, Harbin Institute of Technology, Harbin, 150001, China
Lan_p@sina.com

**Abstract-The aim of this study is to establish a general transformation matrix between B-spline surfaces and ANCF surface elements. This study is a further study of the conversion between the ANCF and B-spline surfaces. In this paper, a general transformation matrix between the $i \times j(i, j \leq 3, i, j \in N^+)$ Bezier surfaces and ANCF surface element is established. This general transformation matrix essentially describes the linear relationship between ANCF and $i \times j(i, j \leq 3, i, j \in N^+)$ Bezier surfaces. Moreover, the general transformation matrix can help to improve the efficiency of the process to transfer the distorted configuration in the CAA back to the CAD, an urgent requirement in engineering practice. In addition, a special Bezier surface control polygon is given in this study. The Bezier surface described with this control polygon can be converted to an ANCF surface element with fewer d.o.f.. And the converted ANCF surface element with 36 d.o.f. was once addressed by Dufva and Shabana. So the special control polygon can be regarded as the geometric condition in conversion to an ANCF surface element with 36 d.o.f.. Based on the fact that a B-spline surface can be seen as a set of Bezier surfaces connected together, the method to establish a general transformation matrix between the ANCF and lower-order $i \times j(i + j < 6, i, j = 1,2,3)$ B-spline surfaces is given. Specially, the general transformation is not in a recursive form, but in a simplified form.**

## 1. INTRODUCTION

Recently, the integration of computer aided design and analysis (ICADA) has become a heated issue. The methods to represent geometries in CAD and CAA software are different essentially [1,2]. In computational geometry and CAD, the typical methods to describe complex geometries are Bezier, B-spline and NURBS representations [1]. But due to the employment of rotations as coordinates in traditional beam, plate and shell elements, these elements cannot described exactly in CAD software. However, the successful conversion between the ANCF and NURBS geometry representation indicates that we can construct a finite element mesh based on the ANCF directly from NURBS [3]. The ANCF finite element does not employ rotations as nodal coordinates, but employ absolute position and gradient coordinates instead [4,5]. This advantage helps to establish the relationship between the ANCF nodal position and gradient coordinates and the control points used in the NURBS. So it facilitates the process of ICADA.

According to the published papers [1,2], the transformation matrix between B-spline curves and ANCF cable elements as well as the transformation matrix between ANCF and B-spline surfaces have be developed and the linear transformation established preserves the ANCF finite element desirable features without changing reference configuration created by CAD. However, there is not a general transformation established between ANCF and Bezier and B-spline surfaces. Moreover, the usual process to convert a lower-order $i \times j(i + j < 6, i, j = 1,2,3)$ B-spline surface to ANCF surface element is (1) ascend the order to the bicubic order. (2) Convert the bicubic Bezier or B-spline surface to the ANCF surface element.

The goal of this study is to establish a general transformation matrix between the ANCF and lower-order Bezier and B-spline surface. It can help to convert a lower-order Bezier and B-spline surfaces to the corresponding ANCF surface element directly, without ascending the order. Specially, one of the advantages of the general transformation matrix between B-spline and ANCF surfaces is that the matrix is given in a simplified form, not a recursive form. In addition, the paper recommends a special Bezier surface control polygon, which can lead to an ANCF finite surface element of fewer degrees of freedom after conversion.

## 2. GENERAL TRANSFORMATION MATRIX BETWEEN ANCF AND BEZIER SURFACES

In the Bezier representation, a Bezier tensor product surface is defined as:

$$\mathbf{p}(u,v)=\sum_{i=0}^{m}\sum_{j=0}^{n}\mathbf{b}_{i,j}B_{i,m}(u)B_{j,n}(v), 0<u,v<1 \quad (1)$$

where, the coefficients $\mathbf{b}_{i,j}(i=0,1,...,m; j=0,1,...,n)$ are the control points. $B_{i,m}(u)$ and $B_{j,n}(v)$ are the basis functions[6].

In the case of the ANCF surface element, the global position rector $\mathbf{r}(u,v)$ of an arbitrary point on the element can be defined using the element shape functions and the nodal coordinate vector:

$$\mathbf{r}(u,v)=\mathbf{S}(u,v)\mathbf{e}(t) \quad (2)$$

where, the matrix of global shape functions are as follows:

$$\mathbf{S}=\left[S_{11}\mathbf{I}, S_{12}\mathbf{I}, S_{13}\mathbf{I}, S_{14}\mathbf{I}; \cdots; S_{41}\mathbf{I}, S_{42}\mathbf{I}, S_{43}\mathbf{I}, S_{44}\mathbf{I}\right] \quad (3)$$

where,

$$S_{ij}=s_i(u,a)s_j(v,b) \quad (4)$$

in which, a is the assumed element length and b is the assumed element width. In Eq. (3), $\mathbf{I}$ refers to the $3\times3$ identity matrix and shape functions are

$$\left.\begin{array}{ll} s_1(p,l)=1-3\lambda^2+2\lambda^3, & s_3(p,l)=3\lambda^2-2\lambda^3 \\ s_2(p,l)=l(\lambda-2\lambda^2+\lambda^3), & s_4(p,l)=l(\lambda^3-\lambda^2) \end{array}\right\} \quad (5)$$

where,

$$\begin{cases} \lambda=\xi=\dfrac{u}{a}, \; if \; p=u, l=a \\ \lambda=\eta=\dfrac{v}{b}, \; if \; p=v, l=b \end{cases} \quad (6)$$

The vector of absolute nodal coordinates referred in Eq. (2) is as follows:

$$\begin{aligned}\mathbf{e}=\Big[&\mathbf{r}_{00}^{00}, \mathbf{r}_{00}^{10}, \mathbf{r}_{a0}^{00}, \mathbf{r}_{a0}^{10}, \mathbf{r}_{00}^{01}, \mathbf{r}_{00}^{11}, \mathbf{r}_{a0}^{01}, \mathbf{r}_{a0}^{11}, \\ &\mathbf{r}_{0b}^{00}, \mathbf{r}_{0b}^{10}, \mathbf{r}_{ab}^{00}, \mathbf{r}_{ab}^{10}, \mathbf{r}_{0b}^{01}, \mathbf{r}_{0b}^{11}, \mathbf{r}_{ab}^{01}, \mathbf{r}_{ab}^{11}\Big]^T\end{aligned} \quad (7)$$

where, each element $\mathbf{r}_{uv}^{ij}$ represents a row vector, which has the expression:

$$\mathbf{r}_{uv}^{ij}=\frac{\partial^{i+j}\mathbf{r}}{\partial \mathrm{u}^i \partial \mathrm{v}^j} \quad (8)$$

in which, if $i=j=0$, it represents displacement vectors, if $i+j=1$, it represents some slope vector and if $i+j=2$, it represents second-order slope vectors[7].

From Eq. (1) and Eq. (3) ~ Eq. (8), we obtain that there is a general linear transformation matrix $\mathbf{T}$ which can convert the $i\times j(i,j\le 3, i,j\in N^+)$ Bezier surface to its corresponding ANCF finite surface element, and one can get the expressions of nodal coordinates from Eq. (1). Eq. (9) gives the conversion with the transformation matrix $\mathbf{T}$:

$$\mathbf{e}=\mathbf{T}_u\mathbf{b}=\begin{bmatrix}\mathbf{T}_{u1} & \mathbf{0} \\ \mathbf{0} & \mathbf{T}_{u2}\end{bmatrix}\mathbf{b} \quad (9)$$

where,

$$
\mathbf{T}_{u1}=\begin{bmatrix}
\mathbf{I} & & & & & & & \\
-\frac{i}{a}\mathbf{I} & \frac{i}{a}\mathbf{I} & & & & \mathbf{0} & & \\
0 & 0 & \mathbf{I} & & & & & \\
0 & 0 & \frac{i}{a}\mathbf{I} & -\frac{i}{a}\mathbf{I} & & & & \\
-\frac{j}{b}\mathbf{I} & 0 & 0 & 0 & \frac{j}{b}\mathbf{I} & & & \\
\frac{ij}{ab}\mathbf{I} & -\frac{ij}{ab}\mathbf{I} & 0 & 0 & -\frac{ij}{ab}\mathbf{I} & \frac{ij}{ab}\mathbf{I} & & \\
0 & 0 & -\frac{j}{b}\mathbf{I} & 0 & 0 & 0 & \frac{j}{b}\mathbf{I} & \\
0 & 0 & -\frac{ij}{ab}\mathbf{I} & \frac{ij}{ab}\mathbf{I} & 0 & 0 & \frac{ij}{ab}\mathbf{I} & -\frac{ij}{ab}\mathbf{I}
\end{bmatrix} \tag{10}
$$

$$
\mathbf{T}_{u2}=\begin{bmatrix}
\mathbf{I} & & & & & & & \\
-\frac{i}{a}\mathbf{I} & \frac{i}{a}\mathbf{I} & & & & \mathbf{0} & & \\
0 & 0 & \mathbf{I} & & & & & \\
0 & 0 & \frac{i}{a}\mathbf{I} & -\frac{i}{a}\mathbf{I} & & & & \\
\frac{j}{b}\mathbf{I} & 0 & 0 & 0 & -\frac{j}{b}\mathbf{I} & & & \\
-\frac{ij}{ab}\mathbf{I} & \frac{ij}{ab}\mathbf{I} & 0 & 0 & \frac{ij}{ab}\mathbf{I} & -\frac{ij}{ab}\mathbf{I} & & \\
0 & 0 & \frac{j}{b}\mathbf{I} & 0 & 0 & 0 & -\frac{j}{b}\mathbf{I} & \\
0 & 0 & \frac{ij}{ab}\mathbf{I} & -\frac{ij}{ab}\mathbf{I} & 0 & 0 & -\frac{ij}{ab}\mathbf{I} & \frac{ij}{ab}\mathbf{I}
\end{bmatrix} \tag{11}
$$

Using the general linear transformation matrix $\mathbf{T}$, we take a cubic × quadratic Bezier surface for numerical example. From the four figures below, we can obtain the linear relationship between the original Bezier control points and the converted ANCF nodal coordinates.

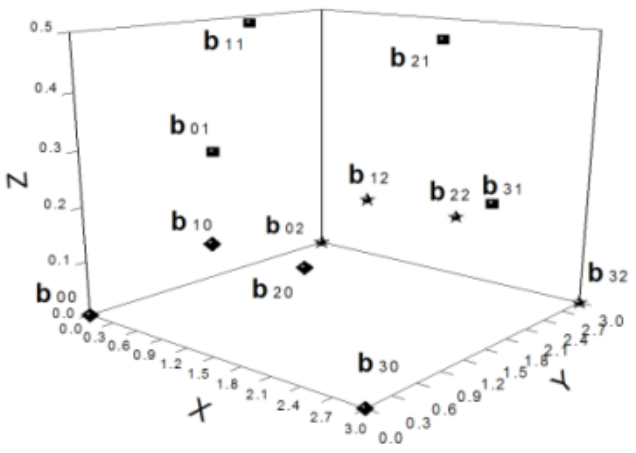

Fig.1 the original bicubic Bezier surface control points

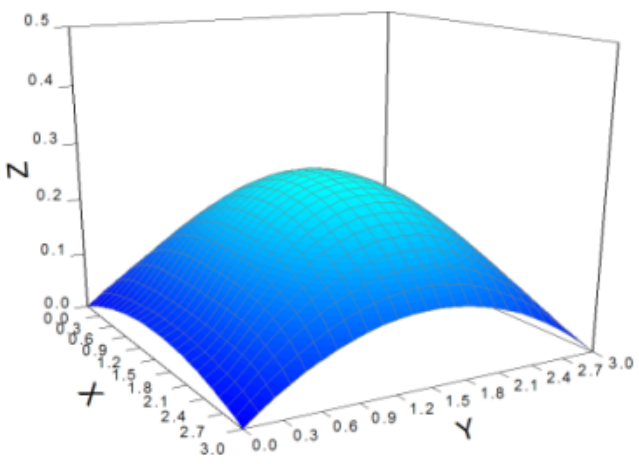

Fig.2 the original cubic × quadratic Bezier surface

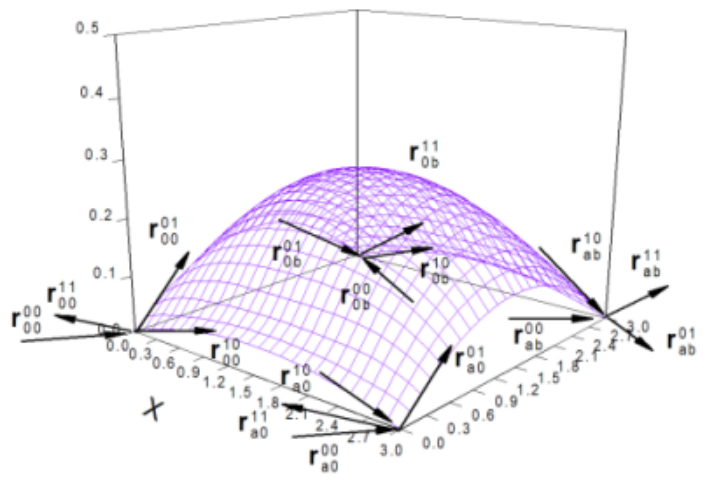

Fig.3 the converted ANCF finite surface element

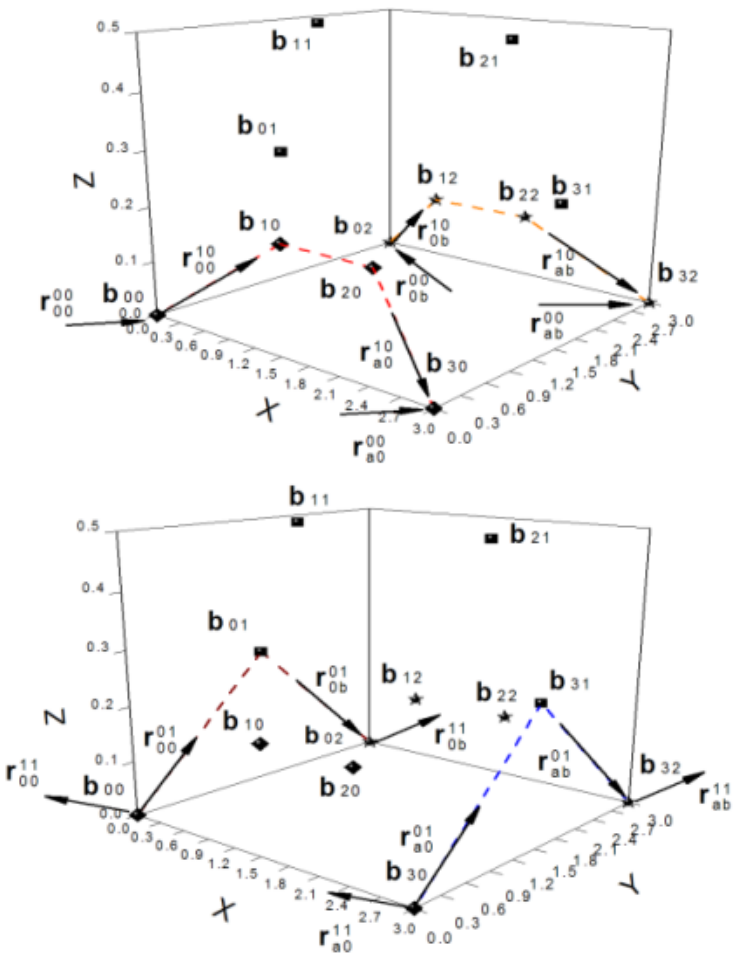

Fig.4 the linear relationship between control points and nodal coordinates

Usually, the process of conversion between ANCF and Bezier surfaces is as follows: 1. Transfer a lower-order Bezier surfaces to a bicubic one by increasing the number of control points. (See Fig.5 and Fig.6) 2. Convert the bicubic Bezier surface to a cubic ANCF surface element. However, the transformation matrix we obtained can help to improve the efficiency of conversion by simplify the conversion to one step and exhibits a general form.

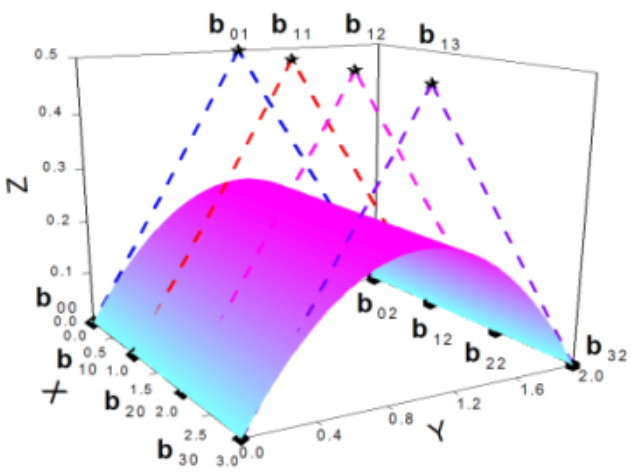

Fig.5 a cubic × quadratic Bezier surface with 12 old control points

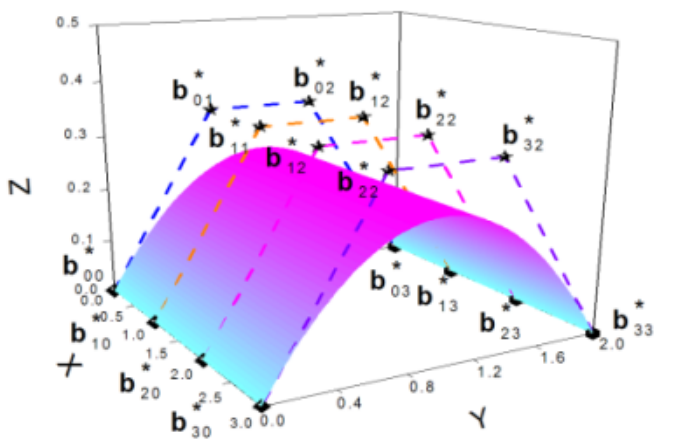

Fig.6 a bicubic Bezier surface with 16 new control points

In Fig.3, the converted nodal coordinates are not independent, because the number of control points of a lower-order Bezier surface is 12, while the number of nodal coordinates of the converted ANCF element is 16. Therefore, $\mathbf{r}_{uv}^{ij}$ only possesses 12 independent nodal coordinates. The other 4 nodal coordinates can be represented by the 12 independent ones. Using the Eq. (9) ~ Eq. (11) and Eq. (1), the relationship between the converted nodal coordinates can be obtained:

$$\begin{cases} \mathbf{r}_{0b}^{01} = \dfrac{2}{b}\mathbf{r}_{0b}^{00} - \dfrac{2}{b}\mathbf{r}_{00}^{00} - \mathbf{r}_{00}^{01} \\ \mathbf{r}_{0b}^{11} = \dfrac{2}{b}\mathbf{r}_{0b}^{10} - \dfrac{2}{b}\mathbf{r}_{00}^{10} - \mathbf{r}_{00}^{11} \\ \mathbf{r}_{ab}^{01} = \dfrac{2}{b}\mathbf{r}_{ab}^{00} - \dfrac{2}{b}\mathbf{r}_{a0}^{00} - \mathbf{r}_{a0}^{01} \\ \mathbf{r}_{ab}^{11} = \dfrac{2}{b}\mathbf{r}_{ab}^{10} - \dfrac{2}{b}\mathbf{r}_{00}^{01} - \mathbf{r}_{a0}^{11} \end{cases} \tag{12}$$

Therefore, we know that if the nodal coordinates of a cubic ANCF surface element meet the condition Eq. (12) and when we convert the cubic ANCF surface to a Bezier surface, we will obtain a lower-order Bezier surface. The reversed process will be helpful in the integration of CAA and CAD.

In the current engineering practice, we are usually acquired to transfer the figures describing the structure after distortion in CAA engineering software to the CAD software in order to describe the distorted configuration. Upon the requirement, it is usually that, only a few part of the structure is distorted while the rest of the configuration of the structure is not distorted or the distortion can be ignored. And the undistorted structure is represented by the lower-order Bezier representation originally. Under this circumstance, we are not required to convert the ANCF to a bicubic Bezier representation, but convert it to a lower-order one directly. This process will improve the efficiency of the conversion.

The general transformation matrix to realize the reversed process is obtained as follows:

$$\mathbf{T}_u^{-1} = \begin{bmatrix} \mathbf{T}_{u1}^{-1} & \mathbf{0} \\ \mathbf{0} & \mathbf{T}_{u2}^{-1} \end{bmatrix} \tag{13}$$

where,

$$\mathbf{T}_{u1}^{-1} = \begin{bmatrix} \mathbf{I} & & & & & & & \\ \mathbf{I} & \frac{a}{i}\mathbf{I} & & & & \mathbf{0} & & \\ 0 & 0 & \mathbf{I} & & & & & \\ 0 & 0 & \mathbf{I} & -\frac{a}{i}\mathbf{I} & & & & \\ \mathbf{I} & 0 & 0 & 0 & \frac{b}{j}\mathbf{I} & & & \\ \mathbf{I} & \frac{a}{i}\mathbf{I} & 0 & 0 & \frac{b}{j}\mathbf{I} & \frac{ab}{ij}\mathbf{I} & & \\ 0 & 0 & \mathbf{I} & 0 & 0 & 0 & \frac{b}{j}\mathbf{I} & \\ 0 & 0 & \mathbf{I} & -\frac{a}{i}\mathbf{I} & 0 & 0 & \frac{b}{j}\mathbf{I} & -\frac{ab}{ij}\mathbf{I} \end{bmatrix} \tag{14}$$

$$\mathbf{T}_{u2}^{-1} = \begin{bmatrix} \mathbf{I} & & & & & & & \\ \mathbf{I} & \frac{a}{i}\mathbf{I} & & & & \mathbf{0} & & \\ 0 & 0 & \mathbf{I} & & & & & \\ 0 & 0 & \mathbf{I} & -\frac{a}{i}\mathbf{I} & & & & \\ \mathbf{I} & 0 & 0 & 0 & -\frac{b}{j}\mathbf{I} & & & \\ \mathbf{I} & \frac{a}{i}\mathbf{I} & 0 & 0 & -\frac{b}{j}\mathbf{I} & -\frac{ab}{ij}\mathbf{I} & & \\ 0 & 0 & \mathbf{I} & 0 & 0 & 0 & -\frac{b}{j}\mathbf{I} & \\ 0 & 0 & \mathbf{I} & -\frac{a}{i}\mathbf{I} & 0 & 0 & -\frac{b}{j}\mathbf{I} & \frac{ab}{ij}\mathbf{I} \end{bmatrix} \tag{15}$$

## 3. CONVERTED ANCF SURFACE ELEMENT WITH FEWER D.O.F.

As an alternative to the 48 degree-of-freedom element, Dufva and Shabana addressed an element by omitting the second order gradient vectors from the generalized coordinates. The exclusion of the second order gradients from the nodal coordinates has some advantages. Eg. Reduced d.o.f. [4].

According to the Eq. (9), if we define the relationship between the 16 control points of the original bicubic Bezier surface as follows:

$$\begin{cases}\mathbf{b}_{0,0}+\mathbf{b}_{1,1}=\mathbf{b}_{1,0}+\mathbf{b}_{0,1}\\ \mathbf{b}_{0,3}+\mathbf{b}_{1,2}=\mathbf{b}_{1,3}+\mathbf{b}_{1,0}\\ \mathbf{b}_{3,0}+\mathbf{b}_{2,1}=\mathbf{b}_{2,0}+\mathbf{b}_{3,1}\\ \mathbf{b}_{3,3}+\mathbf{b}_{2,2}=\mathbf{b}_{2,3}+\mathbf{b}_{3,2}\end{cases}\tag{16}$$

then with Eq. (9), we can obtain $\mathbf{r}_{uv}^{11}=0$.

It means that the quadrilaterals $\mathbf{b}_{0,0}\mathbf{b}_{1,0}\mathbf{b}_{1,1}\mathbf{b}_{0,1}$, $\mathbf{b}_{0,3}\mathbf{b}_{1,3}\mathbf{b}_{1,2}\mathbf{b}_{1,0}$, $\mathbf{b}_{3,0}\mathbf{b}_{2,0}\mathbf{b}_{2,1}\mathbf{b}_{3,1}$, $\mathbf{b}_{3,3}\mathbf{b}_{2,3}\mathbf{b}_{2,2}\mathbf{b}_{3,2}$ are four parallelograms. And the nodal coordinate vectors of the reduced-order ANCF finite surface element will be defined as follows:

$$\mathbf{e}=\left[\mathbf{r}_{00}^{00},\mathbf{r}_{00}^{10},\mathbf{r}_{a0}^{00},\mathbf{r}_{a0}^{10},\mathbf{r}_{00}^{01},\mathbf{r}_{a0}^{01},\mathbf{r}_{0b}^{00},\mathbf{r}_{0b}^{10},\mathbf{r}_{ab}^{00},\mathbf{r}_{ab}^{10},\mathbf{r}_{0b}^{01},\mathbf{r}_{ab}^{01}\right]^T\tag{17}$$

The new shape function matrix $\mathbf{S}^*$ will be defined as

$$\begin{aligned}\mathbf{S}^*=[&S_{11}\mathbf{I},S_{12}\mathbf{I},S_{13}\mathbf{I},S_{14}\mathbf{I};S_{21}\mathbf{I},S_{23}\mathbf{I};\\&S_{31}\mathbf{I},S_{32}\mathbf{I},S_{33}\mathbf{I},S_{34}\mathbf{I};S_{41}\mathbf{I},S_{43}\mathbf{I}]\end{aligned}\tag{18}$$

where, $S_{i,j}$ refers to Eq. (4) ~ Eq. (6).

## 4. General transformation matrix between ANCF and B-spline surfaces

A $k\times l$ B-spline tensor product surface is defined as follows [6]:

$$\mathbf{p}(u,v)=\sum_{i=0}^{m}\sum_{j=0}^{n}\mathbf{d}_{i,j}N_{i,k}(u)B_{j,l}(v),\ u_k\le u\le u_{m+1},\ v_l\le v\le v_{n+1}\tag{19}$$

In Eq. (19), $\mathbf{d}_{i,j}(i=0,1,\cdots,m;j=0,1,\cdots,n)$ are the control points, $N_{i,k}(u)$, $B_{j,l}(v)$ are B-spline basis functions, which are defined in a recursive form:

$$\begin{cases}N_{i,0}(u)=\begin{cases}1 & if\ u_i\le u\le u_{i+1}\\ 0 & otherwise\end{cases}\\ N_{i,k}(u)=\dfrac{u-u_i}{u_{i+k}-u_i}N_{i,j-1}(u)+\dfrac{u_{i+k+1}-u}{u_{i+k+1}-u_{i+1}}N_{i+1,k-1}(u)\\ define\ \dfrac{0}{0}=0\end{cases}\tag{20}$$

$$\begin{cases}B_{j,0}(v)=\begin{cases}1 & if\ v_j\le v\le v_{j+1}\\ 0 & otherwise\end{cases}\\ B_{j,l}(v)=\dfrac{v-v_i}{v_{j+l}-v_j}B_{j,l-1}(v)+\dfrac{v_{j+l+1}-v}{v_{j+l+1}-v_{j+1}}B_{j+1,l-1}(v)\\ define\ \dfrac{0}{0}=0\end{cases}\tag{21}$$

where, $u_i(i=0,1,2,\dots,m+k+1)$ and $v_j(j=0,1,2,\dots,n+l+1)$ are called the knots. $\mathbf{U}=\{u_0,u_1,\cdots,u_{m+k+1}\}$ and $\mathbf{V}=\{v_0,v_1,\cdots,v_{n+l+1}\}$ are called the knot vectors.

Suppose there is a non-zero area $u_e\le u\le u_{e+1}$, $v_f\le v\le v_{f+1}$, the expression of a bicubic B-spline surface defined in this area can be written as:

$$\mathbf{p}(u,v)=\sum_{i=e-3}^{e}\sum_{j=f-3}^{f}\mathbf{d}_{i,j}N_{i,3}(u)B_{j,3}(v)\tag{22}$$

The knot vectors along the two directions of parameter $u$ and $v$ are defined as follows:

$$U_e=\{u_{e-2},u_{e-1},u_e,u_{e+1},u_{e+2},u_{e+3}\}\tag{23}$$

$$V_f=\{v_{f-2},v_{f-1},v_f,v_{f+1},v_{f+2},v_{f+3}\}\tag{24}$$

Using Eq. (12) ~ Eq. (14), one can get the expression of the bicubic B-spline surface:

$$\begin{cases}K_{\alpha-3,3}(\lambda)=\dfrac{F_1^3}{H_{1,-2}H_{1,-1}H_{1,0}}\\ K_{\alpha-2,3}(\lambda)=\dfrac{F_1^2G_{-2}}{H_{1,-2}H_{1,-1}H_{1,0}}+\dfrac{F_1F_2G_{-1}}{H_{2,-1}H_{1,-1}H_{1,0}}+\dfrac{F_2^2G_0}{H_{2,-1}H_{2,0}H_{1,0}}\\ K_{\alpha-1,3}(\lambda)=\dfrac{F_1G_{-1}^2}{H_{2,-1}H_{1,-1}H_{1,0}}+\dfrac{F_2G_0G_{-1}}{H_{2,-1}H_{2,0}H_{1,0}}+\dfrac{F_3G_0^2}{H_{3,0}H_{2,0}H_{1,0}}\\ K_{\alpha,3}(\lambda)=\dfrac{G_0^3}{H_{3,0}H_{2,0}H_{1,0}}\end{cases}\tag{25}$$

where,

$$\begin{cases}F_\beta=\lambda_{\alpha+\beta}-\lambda\\ G_m=\lambda-\lambda_{\alpha+\beta}\\ H_{\beta,\gamma}=\lambda_{\alpha+\beta}-\lambda_{\alpha+\gamma}\end{cases}\tag{26}$$

and

$$\begin{cases} K = N, \alpha = e, \beta = m, \gamma = n & if\, \lambda = u \\ K = B, \alpha = f, \beta = p, \gamma = q & if\, \lambda = v \end{cases} \quad (27)$$

Using Eq. (25) ~ Eq. (27), it can be obtained that $K_{\alpha-3,3}(\lambda_{i+1}) = 0$ and $K_{\alpha,3}(\lambda_i) = 0$.

Using Eq. (19), the nodal coordinates of corresponding ANCF converted from a bicubic B-spline surface can be obtained:

$$\begin{cases} \mathbf{r}_{uv}^{00} = \mathbf{p}(u_a, v_b) \\ \mathbf{r}_{uv}^{10} = \dfrac{\partial \mathbf{p}(u,v)}{\partial u} \dfrac{\partial u}{\partial \xi} \dfrac{\partial \xi}{\partial x} \Big|_{u=u_a, v=v_b} \\ \mathbf{r}_{uv}^{01} = \dfrac{\partial \mathbf{p}(u,v)}{\partial v} \dfrac{\partial v}{\partial \eta} \dfrac{\partial \eta}{\partial y} \Big|_{u=u_a, v=v_b} \\ \mathbf{r}_{uv}^{11} = \dfrac{\partial \mathbf{p}(u,v)}{\partial u \partial v} \dfrac{\partial u}{\partial \xi} \dfrac{\partial \xi}{\partial x} \dfrac{\partial v}{\partial \eta} \dfrac{\partial \eta}{\partial y} \Big|_{u=u_a, v=v_b} \end{cases} \quad (28)$$

where, $a = i, i+1$, $b = j, j+1$.

The relationship between the parameter $x, y$ referred in the ANCF finite surface element and the assumed length $a$ and width $b$ of the bicubic B-spline surface is as follows:

$$\begin{cases} x = \xi a \\ y = \eta b \end{cases} \quad (29)$$

The relationship between the parameters $\xi, \eta$ referred in the ANCF finite surface element and the parameters $u, v$ of B-spline parametric surface is as follows:

$$\begin{cases} \xi = (u - u_i)/(u_{i+1} - u_i) \\ \eta = (v - v_i)/(v_{i+1} - v_i) \end{cases} \quad (30)$$

Using Eq. (22) ~ Eq. (30), the 16 nodal coordinates of the converted ANCF finite surface element can be obtained: (refer to appendix A)

Therefore, from appendix A, the transformation matrix from a bicubic B-spline surface to its corresponding ANCF finite surface element can be obtained as follows:

$$\mathbf{r}_{uv}^{ij} = \left[\mathbf{T}^{Bs(3\times3)}\right] \mathbf{d}^{Bs(3\times3)} = \begin{bmatrix} \mathbf{T}_{00}^{Bs} & \mathbf{T}_{01}^{Bs} & \mathbf{T}_{02}^{Bs} & \mathbf{T}_{03}^{Bs} \\ \mathbf{T}_{10}^{Bs} & \mathbf{T}_{11}^{Bs} & \mathbf{T}_{12}^{Bs} & \mathbf{T}_{13}^{Bs} \\ \mathbf{T}_{20}^{Bs} & \mathbf{T}_{21}^{Bs} & \mathbf{T}_{22}^{Bs} & \mathbf{T}_{23}^{Bs} \\ \mathbf{T}_{30}^{Bs} & \mathbf{T}_{31}^{Bs} & \mathbf{T}_{32}^{Bs} & \mathbf{T}_{33}^{Bs} \end{bmatrix} \mathbf{d}^{Bs(3\times3)} \quad (31)$$

in which, the elements in transformation matrix refer to appendix B.

$$\mathbf{d}^{Bs(3\times3)} = \left[\mathbf{d}_{e-3,f-3}, \mathbf{d}_{e-3,f-2}, \mathbf{d}_{e-3,f-1}, \mathbf{d}_{e-3,f}, \mathbf{d}_{e-2,f-3}, \mathbf{d}_{e-2,f-2}, \mathbf{d}_{e-2,f-1}, \mathbf{d}_{e-2,f}, \mathbf{d}_{e-1,f-3}, \mathbf{d}_{e-1,f-2}, \mathbf{d}_{e-1,f-1}, \mathbf{d}_{e-1,f}, \mathbf{d}_{e,f-3}, \mathbf{d}_{e,f-2}, \mathbf{d}_{e,f-1}, \mathbf{d}_{e,f}\right] \quad (32)$$

Similarly, a cubic $\times$ quadratic B-spline surface can also be converted to an ANCF finite surface element exactly. The knot vectors along the two directions of parameter $u$ and $v$ are defined as follows:

$$U_e = \{u_{e-2}, u_{e-1}, u_e, u_{e+1}, u_{e+2}, u_{e+3}\} \quad (33)$$

$$V_f = \{v_{f-1}, v_f, v_{f+1}, v_{f+2}\} \quad (34)$$

Using Eq. (19) and Eq. (20), the basis function $B_{j,2}(v)$ and the basis function $N_{i,2}(u)$ used in the conversion from the quadratic × cubic B-spline surface to ANCF finite surface element can be obtained:

$$\begin{cases} K_{\alpha-2,2}(\lambda) = \dfrac{F_1^2}{H_{1,-1}} \\ K_{\alpha-1,2}(\lambda) = \dfrac{F_1 G_{-1}}{H_{1,0} H_{1,-1}} + \dfrac{F_2 G_0}{H_{2,0} H_{1,0}} \\ K_{\alpha,2}(\lambda) = \dfrac{G_0^2}{H_{2,0}} \end{cases} \quad (35)$$

Moreover, one can obtain from Eq. (36) that

$$\begin{cases} K_{\alpha-2,2}(\lambda_i)=\dfrac{H_{1,0}^2}{H_{1,-1}},\ K_{\alpha-1,2}(\lambda_i)=\dfrac{H_{0,-1}}{H_{1,-1}} \\ K'_{\alpha-2,2}(\lambda_i)=-\dfrac{2H_{1,0}}{H_{1,-1}},K'_{\alpha-1,2}(\lambda_i)=\dfrac{2}{H_{1,-1}} \\ K_{\alpha-1,2}(\lambda_{i+1})=\dfrac{H_{2,1}}{H_{2,0}},K_{\alpha,2}(\lambda_{i+1})=\dfrac{H_{1,0}^2}{H_{2,0}} \\ K'_{\alpha-1,2}(\lambda_{i+1})=-\dfrac{2}{H_{2,0}},K'_{\alpha,2}(\lambda_{i+1})=\dfrac{2H_{1,0}}{H_{2,0}} \end{cases} \tag{36}$$

Therefore, using Eq. (22), Eq. (28) ~ Eq. (30), appendix A, and Eq. (35) ~ Eq. (36), the transformation matrix used to convert a cubic × quadratic B-spline surface to its corresponding ANCF finite surface element can be obtained

From Eq. (36) ,we know that $K_{\alpha-2,2}(\lambda_{i+1})=0$ and $K_{\alpha,2}(\lambda_i)=0$. It means that for the segment of a parametric surface, one can suppose the four vertices of it are $\mathbf{p}(u_i,v_j),\mathbf{p}(u_i,v_{j+1}),\mathbf{p}(u_{i+1},v_j)$ and $\mathbf{p}(u_{i+1},v_{j+1})$. These four vertices cannot be influenced by the coordinate of $\mathbf{d}_{i,j},\mathbf{d}_{i,j-2},\mathbf{d}_{i-3,j}$ and $\mathbf{d}_{i-3,j-2}$, respectively. And when $k=1$ or $l=1$, the basis functions $K_{\alpha-1,1}(\lambda)=F_1$ and $K_{\alpha,1}(\lambda)=G_0$ will be used.

## 5. CONCLUSION

This paper established a general transformation matrix between the ANCF and lower-order $i\times j(i+j<6\,,i,j\le 3,i,j\in N^+)$ Bezier and B-spline surfaces. The general transformation matrix is a generalization of the transformation matrix between the bicubic Bezier and B-spline surface and ANCF surface element in the published paper.

The special Bezier surface control polygon proposed in this study leads to an ANCF finite surface element converted of 36-d.o.f. This converted thin plate has its advantages in combination between the converted ANCF element because the elimination of ANCF nodal coordinates $\mathbf{r}_{00}^{11},\mathbf{r}_{a0}^{11},\mathbf{r}_{0b}^{11}$ and $\mathbf{r}_{ab}^{11}$.

As demonstrated in this study, using the general inverse transformation matrix, the converted Bezier and B-spline surface will be a lower-order one directly if the dependence of original ANCF nodal coordinates meets the conditions referred. This process enhances the efficiency of I-CAD-A and meets the urgent engineering requirement.

## Acknowledgements

This research is supported by National Natural Science Foundation of China (Grant No. 11172076 ) and by Science and Technology Innovation Talent Foundation of Harbin (2012RFLXG020).

## 6. REFERENCE

1. Lan, P., Shabana, A.A.: Integration of b-spline geometry and ANCF finite element analysis. Nonlinear Dyn. 61(1–2), 193–206 (2010)
2. Aki Mikkola, Ahmed A. Shabana, Cristina Sanchez-Rebollo, Jesus R. Jimenez-Octavio: Comparison between ANCF and B-spline surfaces.Multibody Syst Dyn. DOI 10.1007/s11044-013-9353-z (2013)
3. Graham G.Sanborn, Ahmed A.Shabana: On the integration of computer aided design and analysis using the finite element absolute nodal coordinate. Multibody Syst Dyn 22: 181-197(2009)
4. K Dufva1 and Ahmed A. Shabana; Analysis of thin plate structures using theabsolute nodal coordinate formulation :Proc. IMechE Vol. 219 Part K: J. Multi-body Dynamics (2005)
5. Lan, P., Shabana: Rational finite elements and flexible body dynamics. Journal of Vibration and Acoustics, Transactions of the ASME, v 132, n 4, p 0410071-0410079, August 2010
6. Piegl, L., Tiller, W.: The NURBS Book, 2nd edn. Springer, New York (1997)
7. O.N. DMITROCHENKO and D.YU. POGORELOV: Generalization of Plate Finite Elements for Absolute Nodal Coordinate Formulation. Multibody System Dynamics 10: 17–43(2003)

## Appendix A

$$
\left\{
\begin{aligned}
&\mathbf{r}_{00}^{00} = \mathbf{p}(u_i, v_j) = \sum_{i=e-3}^{e-1} \sum_{j=f-3}^{f-1} \mathbf{d}_{i,j} N_{i,3}(u_i) B_{j,3}(v_j) \\
&\mathbf{r}_{00}^{10} = \frac{\partial \mathbf{p}(u,v)}{\partial u}\Big|_{u=u_i, v=v_j} = \frac{1}{a}\left(u_{i+1}-u_i\right) \sum_{i=e-3}^{e-1} \sum_{j=f-3}^{f-1} \mathbf{d}_{i,j} \left(N_{i,3}\right)_{,u} \big|_{u=u_i} B_{j,3}(v_j) \\
&\mathbf{r}_{00}^{01} = \frac{\partial \mathbf{p}(u,v)}{\partial v}\Big|_{u=u_i, v=v_j} = \frac{1}{b}\left(v_{j+1}-v_j\right) \sum_{i=e-3}^{e-1} \sum_{j=f-3}^{f-1} \mathbf{d}_{i,j} N_{i,3}(u_i) \left(B_{j,3}\right)_{,v} \big|_{v=v_j} \\
&\mathbf{r}_{00}^{11} = \frac{\partial \mathbf{p}(u,v)}{\partial u \partial v}\Big|_{u=u_i, v=v_j} = \frac{1}{a}\frac{1}{b}\left(u_{i+1}-u_i\right)\left(v_{j+1}-v_j\right) \sum_{i=e-3}^{e-1} \sum_{j=f-3}^{f-1} \mathbf{d}_{i,j} \left(N_{i,3}\right)_{,u} \big|_{u=u_i} \left(B_{j,3}\right)_{,v} \big|_{v=v_j} \\
&\mathbf{r}_{a0}^{00} = \mathbf{p}(u_{i+1}, v_j) = \sum_{i=e-2}^{e} \sum_{j=f-3}^{f-1} \mathbf{d}_{i,j} N_{i,3}(u_{i+1}) B_{j,3}(v_j) \\
&\mathbf{r}_{a0}^{10} = \frac{\partial \mathbf{p}(u,v)}{\partial u}\Big|_{u=u_{i+1}, v=v_j} = \frac{1}{a}\left(u_{i+1}-u_i\right) \sum_{i=e-2}^{e} \sum_{j=f-3}^{f-1} \mathbf{d}_{i,j} \left(N_{i,3}\right)_{,u} \big|_{u=u_{i+1}} B_{j,3}(v_j) \\
&\mathbf{r}_{a0}^{01} = \frac{\partial \mathbf{p}(u,v)}{\partial v}\Big|_{u=u_{i+1}, v=v_j} = \frac{1}{b}\left(v_{j+1}-v_j\right) \sum_{i=e-3}^{e} \sum_{j=f-3}^{f-1} \mathbf{d}_{i,j} N_{i,3}(u_{i+1}) \left(B_{j,3}\right)_{,v} \big|_{v=v_j} \\
&\mathbf{r}_{a0}^{11} = \frac{\partial \mathbf{p}(u,v)}{\partial u \partial v}\Big|_{u=u_{i+1}, v=v_j} = \frac{1}{a}\frac{1}{b}\left(u_{i+1}-u_i\right)\left(v_{j+1}-v_j\right) \sum_{i=e-3}^{e} \sum_{j=f-3}^{f-1} \mathbf{d}_{i,j} \left(N_{i,3}\right)_{,u} \big|_{u=u_{i+1}} \left(B_{j,3}\right)_{,v} \big|_{v=v_j} \\
&\mathbf{r}_{0b}^{00} = \mathbf{p}(u_i, v_{j+1}) = \sum_{i=e-3}^{e-1} \sum_{j=f-2}^{f} \mathbf{d}_{i,j} N_{i,3}(u_i) B_{j,3}(v_{j+1}) \\
&\mathbf{r}_{0b}^{10} = \frac{\partial \mathbf{p}(u,v)}{\partial u}\Big|_{u=u_i, v=v_{j+1}} = \frac{1}{a}\left(u_{i+1}-u_i\right) \sum_{i=e-3}^{e-1} \sum_{j=f-2}^{f} \mathbf{d}_{i,j} \left(N_{i,3}\right)_{,u} \big|_{u=u_i} B_{j,3}(v_{j+1}) \\
&\mathbf{r}_{0b}^{01} = \frac{\partial \mathbf{p}(u,v)}{\partial v}\Big|_{u=u_i, v=v_{j+1}} = \frac{1}{b}\left(v_{j+1}-v_j\right) \sum_{i=e-3}^{e-1} \sum_{j=f-2}^{f} \mathbf{d}_{i,j} N_{i,3}(u_i) \left(B_{j,3}\right)_{,v} \big|_{v=v_{j+1}} \\
&\mathbf{r}_{0b}^{11} = \frac{\partial \mathbf{p}(u,v)}{\partial u \partial v}\Big|_{u=u_i, v=v_{j+1}} = \frac{1}{a}\frac{1}{b}\left(u_{i+1}-u_i\right)\left(v_{j+1}-v_j\right) \sum_{i=e-3}^{e-1} \sum_{j=f-2}^{f} \mathbf{d}_{i,j} \left(N_{i,3}\right)_{,u} \big|_{u=u_i} \left(B_{j,3}\right)_{,v} \big|_{v=v_{j+1}} \\
&\mathbf{r}_{ab}^{00} = \mathbf{p}(u_{i+1}, v_{j+1}) = \sum_{i=e-2}^{e} \sum_{j=f-2}^{f} \mathbf{d}_{i,j} N_{i,3}(u_{i+1}) B_{j,3}(v_{j+1}) \\
&\mathbf{r}_{ab}^{10} = \frac{\partial \mathbf{p}(u,v)}{\partial u}\Big|_{u=u_{i+1}, v=v_{j+1}} = \frac{1}{a}\left(u_{i+1}-u_i\right) \sum_{i=e-2}^{e} \sum_{j=f-2}^{f} \mathbf{d}_{i,j} \left(N_{i,3}\right)_{,u} \big|_{u=u_{i+1}} B_{j,3}(v_{j+1}) \\
&\mathbf{r}_{ab}^{01} = \frac{\partial \mathbf{p}(u,v)}{\partial v}\Big|_{u=u_{i+1}, v=v_{j+1}} = \frac{1}{b}\left(v_{j+1}-v_j\right) \sum_{i=e-2}^{e} \sum_{j=f-2}^{f} \mathbf{d}_{i,j} N_{i,3}(u_{i+1}) \left(B_{j,3}\right)_{,v} \big|_{v=v_{j+1}} \\
&\mathbf{r}_{ab}^{11} = \frac{\partial \mathbf{p}(u,v)}{\partial u \partial v}\Big|_{u=u_{i+1}, v=v_{j+1}} = \frac{1}{a}\frac{1}{b}\left(u_{i+1}-u_i\right)\left(v_{j+1}-v_j\right) \sum_{i=e-2}^{e} \sum_{j=f-2}^{f} \mathbf{d}_{i,j} \left(N_{i,3}\right)_{,u} \big|_{u=u_{i+1}} \left(B_{j,3}\right)_{,v} \big|_{v=v_{j+1}}
\end{aligned}
\right. \tag{37}
$$

where,

$$
\begin{cases}
\theta_3 = K_{i-3,3}(\lambda_\omega) = \dfrac{H_{1,0}^2}{H_{1,-2}H_{1,-1}} \\
\theta_2 = K_{i-2,3}(\lambda_\omega) = \dfrac{H_{0,-2}H_{1,0}}{H_{1,-2}H_{1,-1}} + \dfrac{H_{0,-1}H_{2,0}}{H_{2,-1}H_{1,-1}} \\
\theta_1 = K_{i-1,3}(\lambda_\omega) = \dfrac{H_{0,-1}^2}{H_{2,-1}H_{1,-1}} \\
\theta_{d3} = K'_{i-3,3}(\lambda_\omega) = \dfrac{-3H_{1,0}}{H_{1,-2}H_{1,-1}} \\
\theta_{d2} = K'_{i-2,3}(\lambda_\omega) = \dfrac{H_{1,0} - 2H_{0,-2}}{H_{1,-2}H_{1,-1}} + \dfrac{H_{-1,0} + 2H_{2,0}}{H_{2,-1}H_{1,-1}} \\
\theta_{d1} = K'_{i-1,3}(\lambda_\omega) = \dfrac{3H_{0,-1}}{H_{2,-1}H_{1,-1}} \\
\varphi_2 = K_{i-2,3}(\lambda_{\omega+1}) = \dfrac{H_{2,1}^2}{H_{2,-1}H_{2,0}} \\
\varphi_1 = K_{i-1,3}(\lambda_{\omega+1}) = \dfrac{H_{1,-1}H_{2,1}}{H_{2,-1}H_{2,0}} + \dfrac{H_{3,1}H_{1,0}}{H_{3,0}H_{2,0}} \\
\varphi_0 = K_{i-0,3}(\lambda_{\omega+1}) = \dfrac{H_{1,0}^2}{H_{3,0}H_{2,0}} \\
\varphi_{d2} = K'_{i-2,3}(\lambda_{\omega+1}) = \dfrac{-3H_{2,1}}{H_{2,-1}H_{2,0}} \\
\varphi_{d1} = K'_{i-1,3}(\lambda_{\omega+1}) = \dfrac{H_{2,1} - 2H_{1,-1}}{H_{2,-1}H_{2,0}} + \dfrac{2H_{3,1} - H_{1,0}}{H_{3,0}H_{2,0}} \\
\varphi_{d0} = K'_{i-0,3}(\lambda_{\omega+1}) = \dfrac{3H_{1,0}}{H_{3,0}H_{2,0}}
\end{cases}
\tag{38}
$$

and,

$$
\begin{cases}
K = N, \omega = i, \theta = U^f, \varphi = U^l \quad if\, \lambda = u \\
K = B, \omega = j, \theta = V^f, \varphi = V^l \quad if\, \lambda = v
\end{cases}
\tag{39}
$$

## Appendix B

$$\mathbf{T}_{00}^{Bs} = \begin{bmatrix} U_3^f V_3^f \mathbf{I} & U_3^f V_2^f \mathbf{I} & U_3^f V_1^f \mathbf{I} & \mathbf{0} \\ \psi U_{d3}^f V_3^f \mathbf{I} & \psi U_{d3}^f V_2^f \mathbf{I} & \psi U_{d3}^f V_1^f \mathbf{I} & \mathbf{0} \\ \zeta U_3^f V_{d3}^f \mathbf{I} & \zeta U_3^f V_{d2}^f \mathbf{I} & \zeta U_3^f V_{d2}^f \mathbf{I} & \mathbf{0} \\ \psi\zeta U_{d3}^f V_{d3}^f \mathbf{I} & \psi\zeta U_{d3}^f V_{d2}^f \mathbf{I} & \psi\zeta U_{d3}^f V_{d1}^f \mathbf{I} & \mathbf{0} \end{bmatrix} \quad (40)$$

$$\mathbf{T}_{01}^{Bs} = \begin{bmatrix} U_2^f V_3^f \mathbf{I} & U_2^f V_2^f \mathbf{I} & U_2^f V_1^f \mathbf{I} & \mathbf{0} \\ \psi U_{d2}^f V_3^f \mathbf{I} & \psi U_{d2}^f V_2^f \mathbf{I} & \psi U_{d2}^f V_1^f \mathbf{I} & \mathbf{0} \\ \zeta U_2^f V_{d3}^f \mathbf{I} & \zeta U_2^f V_{d2}^f \mathbf{I} & \zeta U_2^f V_{d2}^f \mathbf{I} & \mathbf{0} \\ \psi\zeta U_{d2}^f V_{d3}^f \mathbf{I} & \psi\zeta U_{d2}^f V_{d2}^f \mathbf{I} & \psi\zeta U_{d2}^f V_{d1}^f \mathbf{I} & \mathbf{0} \end{bmatrix} \quad (41)$$

$$\mathbf{T}_{02}^{Bs} = \begin{bmatrix} U_1^f V_3^f \mathbf{I} & U_1^f V_2^f \mathbf{I} & U_1^f V_1^f \mathbf{I} & \mathbf{0} \\ \psi U_{d1}^f V_3^f \mathbf{I} & \psi U_{d1}^f V_2^f \mathbf{I} & \psi U_{d1}^f V_1^f \mathbf{I} & \mathbf{0} \\ \zeta U_1^f V_{d3}^f \mathbf{I} & \zeta U_1^f V_{d2}^f \mathbf{I} & \zeta U_1^f V_{d1}^f \mathbf{I} & \mathbf{0} \\ \psi\zeta U_{d1}^f V_{d3}^f \mathbf{I} & \psi\zeta U_{d1}^f V_{d2}^f \mathbf{I} & \psi\zeta U_{d1}^f V_{d1}^f \mathbf{I} & \mathbf{0} \end{bmatrix} \quad (42)$$

$$\mathbf{T}_{03}^{Bs} = [\mathbf{0}]_{12\times 12} \quad (43)$$

$$\mathbf{T}_{10}^{Bs} = [\mathbf{0}]_{12\times 12} \quad (44)$$

$$\mathbf{T}_{11}^{Bs} = \begin{bmatrix} U_2^l V_3^f \mathbf{I} & U_2^l V_2^f \mathbf{I} & U_2^l V_1^f \mathbf{I} & \mathbf{0} \\ \psi U_{d2}^l V_3^f \mathbf{I} & \psi U_{d2}^l V_2^f \mathbf{I} & \psi U_{d2}^l V_1^f \mathbf{I} & \mathbf{0} \\ \zeta U_2^l V_{d3}^f \mathbf{I} & \zeta U_2^l V_{d2}^f \mathbf{I} & \zeta U_2^l V_{d2}^f \mathbf{I} & \mathbf{0} \\ \psi\zeta U_{d2}^l V_{d3}^f \mathbf{I} & \psi\zeta U_{d2}^l V_{d2}^f \mathbf{I} & \psi\zeta U_{d2}^l V_{d1}^f \mathbf{I} & \mathbf{0} \end{bmatrix} \quad (45)$$

$$\mathbf{T}_{12}^{Bs} = \begin{bmatrix} U_1^l V_3^f \mathbf{I} & U_1^l V_2^f \mathbf{I} & U_1^l V_1^f \mathbf{I} & \mathbf{0} \\ \psi U_{d1}^l V_3^f \mathbf{I} & \psi U_{d1}^l V_2^f \mathbf{I} & \psi U_{d1}^l V_1^f \mathbf{I} & \mathbf{0} \\ \zeta U_1^l V_{d3}^f \mathbf{I} & \zeta U_1^l V_{d2}^f \mathbf{I} & \zeta U_1^l V_{d1}^f \mathbf{I} & \mathbf{0} \\ \psi\zeta U_{d1}^l V_{d3}^f \mathbf{I} & \psi\zeta U_{d1}^l V_{d2}^f \mathbf{I} & \psi\zeta U_{d1}^l V_{d1}^f \mathbf{I} & \mathbf{0} \end{bmatrix} \quad (46)$$

$$\mathbf{T}_{13}^{Bs} = \begin{bmatrix} U_0^l V_3^f \mathbf{I} & U_0^l V_2^f \mathbf{I} & U_0^l V_1^f \mathbf{I} & \mathbf{0} \\ \psi U_{d0}^l V_3^f \mathbf{I} & \psi U_{d0}^l V_2^f \mathbf{I} & \psi U_{d0}^l V_1^f \mathbf{I} & \mathbf{0} \\ \zeta U_0^l V_{d3}^f \mathbf{I} & \zeta U_0^l V_{d2}^f \mathbf{I} & \zeta U_0^l V_{d1}^f \mathbf{I} & \mathbf{0} \\ \psi\zeta U_{d0}^l V_{d3}^f \mathbf{I} & \psi\zeta U_{d0}^l V_{d2}^f \mathbf{I} & \psi\zeta U_{d0}^l V_{d1}^f \mathbf{I} & \mathbf{0} \end{bmatrix} \quad (47)$$

$$\mathbf{T}_{20}^{Bs} = \begin{bmatrix} \mathbf{0} & U_3^f V_2^l \mathbf{I} & U_3^f V_1^l \mathbf{I} & U_3^f V_0^l \mathbf{I} \\ \mathbf{0} & \psi U_{d3}^f V_2^l \mathbf{I} & \psi U_{d3}^f V_1^l \mathbf{I} & \psi U_{d3}^f V_0^l \mathbf{I} \\ \mathbf{0} & \zeta U_3^f V_{d2}^l \mathbf{I} & \zeta U_3^f V_{d1}^l \mathbf{I} & \zeta U_3^f V_{d0}^l \mathbf{I} \\ \mathbf{0} & \psi\zeta U_{d3}^f V_{d2}^l \mathbf{I} & \psi\zeta U_{d3}^f V_{d1}^l \mathbf{I} & \psi\zeta U_{d3}^f V_{d1}^l \mathbf{I} \end{bmatrix} \quad (48)$$

$$\mathbf{T}_{21}^{Bs} = \begin{bmatrix} \mathbf{0} & U_2^f V_2^l \mathbf{I} & U_2^f V_1^l \mathbf{I} & U_2^f V_0^l \mathbf{I} \\ \mathbf{0} & \psi U_{d2}^f V_2^l \mathbf{I} & \psi U_{d2}^f V_1^l \mathbf{I} & \psi U_{d2}^f V_0^l \mathbf{I} \\ \mathbf{0} & \zeta U_2^f V_{d2}^l \mathbf{I} & \zeta U_2^f V_{d1}^l \mathbf{I} & \zeta U_2^f V_{d0}^l \mathbf{I} \\ \mathbf{0} & \psi\zeta U_{d2}^f V_{d2}^l \mathbf{I} & \psi\zeta U_{d2}^f V_{d1}^l \mathbf{I} & \psi\zeta U_{d2}^f V_{d0}^l \mathbf{I} \end{bmatrix} \quad (49)$$

$$\mathbf{T}_{22}^{Bs} = \begin{bmatrix} \mathbf{0} & U_1^f V_2^l \mathbf{I} & U_1^f V_1^l \mathbf{I} & U_1^f V_0^l \mathbf{I} \\ \mathbf{0} & \psi U_{d1}^f V_2^l \mathbf{I} & \psi U_{d1}^f V_1^l \mathbf{I} & \psi U_{d1}^f V_0^l \mathbf{I} \\ \mathbf{0} & \zeta U_1^f V_{d2}^l \mathbf{I} & \zeta U_1^f V_{d1}^l \mathbf{I} & \zeta U_1^f V_{d0}^l \mathbf{I} \\ \mathbf{0} & \psi\zeta U_{d1}^f V_{d2}^l \mathbf{I} & \psi\zeta U_{d1}^f V_{d1}^l \mathbf{I} & \psi\zeta U_{d1}^f V_{d0}^l \mathbf{I} \end{bmatrix} \quad (50)$$

$$\mathbf{T}_{23}^{Bs} = [\mathbf{0}]_{12\times 12} \quad (51)$$

$$\mathbf{T}_{30}^{Bs} = [\mathbf{0}]_{12\times 12} \quad (52)$$

$$\mathbf{T}_{31}^{Bs} = \begin{bmatrix} \mathbf{0} & U_2^l V_2^l \mathbf{I} & U_2^l V_1^l \mathbf{I} & U_2^l V_0^l \mathbf{I} \\ \mathbf{0} & \psi U_{d2}^l V_2^l \mathbf{I} & \psi U_{d2}^l V_1^l \mathbf{I} & \psi U_{d2}^l V_0^l \mathbf{I} \\ \mathbf{0} & \zeta U_2^l V_{d2}^l \mathbf{I} & \zeta U_2^l V_{d1}^l \mathbf{I} & \zeta U_2^l V_{d0}^l \mathbf{I} \\ \mathbf{0} & \psi\zeta U_{d2}^l V_{d2}^l \mathbf{I} & \psi\zeta U_{d2}^l V_{d1}^l \mathbf{I} & \psi\zeta U_{d2}^l V_{d0}^l \mathbf{I} \end{bmatrix} \quad (53)$$

$$\mathbf{T}_{32}^{Bs} = \begin{bmatrix} \mathbf{0} & U_1^l V_2^l \mathbf{I} & U_1^l V_1^l \mathbf{I} & U_1^l V_0^l \mathbf{I} \\ \mathbf{0} & \psi U_{d1}^l V_2^l \mathbf{I} & \psi U_{d1}^l V_1^l \mathbf{I} & \psi U_{d1}^l V_0^l \mathbf{I} \\ \mathbf{0} & \zeta U_1^l V_{d2}^l \mathbf{I} & \zeta U_1^l V_{d1}^l \mathbf{I} & \zeta U_1^l V_{d0}^l \mathbf{I} \\ \mathbf{0} & \psi\zeta U_{d1}^l V_{d2}^l \mathbf{I} & \psi\zeta U_{d1}^l V_{d1}^l \mathbf{I} & \psi\zeta U_{d1}^l V_{d0}^l \mathbf{I} \end{bmatrix} \quad (54)$$

$$\mathbf{T}_{33}^{Bs} = \begin{bmatrix} \mathbf{0} & U_0^l V_2^l \mathbf{I} & U_0^l V_1^l \mathbf{I} & U_0^l V_0^l \mathbf{I} \\ \mathbf{0} & \psi U_{d0}^l V_2^l \mathbf{I} & \psi U_{d0}^l V_1^l \mathbf{I} & \psi U_{d0}^l V_0^l \mathbf{I} \\ \mathbf{0} & \zeta U_0^l V_{d2}^l \mathbf{I} & \zeta U_0^l V_{d1}^l \mathbf{I} & \zeta U_0^l V_{d0}^l \mathbf{I} \\ \mathbf{0} & \psi\zeta U_{d0}^l V_{d2}^l \mathbf{I} & \psi\zeta U_{d0}^l V_{d1}^l \mathbf{I} & \psi\zeta U_{d0}^l V_{d0}^l \mathbf{I} \end{bmatrix} \quad (55)$$

In Eq. (41) ~ Eq. (56), $\psi = \frac{1}{a}(u_{i+1} - u_i)$ and $\zeta = \frac{1}{b}(v_{j+1} - v_j)$.